\documentclass[a4paper,12pt]{scrartcl}
\makeatletter
\DeclareOldFontCommand{\rm}{\normalfont\rmfamily}{\mathrm}
\DeclareOldFontCommand{\sf}{\normalfont\sffamily}{\mathsf}
\DeclareOldFontCommand{\tt}{\normalfont\ttfamily}{\mathtt}
\DeclareOldFontCommand{\bf}{\normalfont\bfseries}{\mathbf}
\DeclareOldFontCommand{\it}{\normalfont\itshape}{\mathit}
\DeclareOldFontCommand{\sl}{\normalfont\slshape}{\@nomath\sl}
\DeclareOldFontCommand{\sc}{\normalfont\scshape}{\@nomath\sc}
\makeatother
\usepackage[utf8x]{inputenc}
\usepackage[T1]{fontenc}
\usepackage{amsmath}
\usepackage{amsfonts}
\usepackage{slashed}
\usepackage{graphicx}
\usepackage{authblk}

\usepackage{subcaption}
\usepackage{simplewick}
\usepackage{verbatim}
\usepackage{hyperref}
\usepackage[dvipsnames]{xcolor}

\newcommand{\arxiv}[2]{[arXiv:\,\href{http://arxiv.org/abs/#1}{\texttt{#1}} [\texttt{#2}]]}
\newcommand{\arxivold}[1]{[arXiv:\,\href{http://arxiv.org/abs/#1}{\texttt{#1}}\,]}

\usepackage{empheq}

\graphicspath{ {../} }
\title{
  A study of center and chiral symmetry realization in thermal $\mathcal{N}=1$ super Yang-Mills theory using the gradient flow }

\author[1]{Georg Bergner\thanks{georg.bergner@uni-jena.de}}
\author[1]{Camilo L\'opez\thanks{camilo.lopez@uni-jena.de}}
\author[2]{Stefano Piemonte\thanks{stefano.piemonte@ur.de}}

\affil[1]{University of Jena, Institute for Theoretical Physics, 
Max-Wien-Platz 1, D-07743 Jena, Germany}
\affil[2]{University of Regensburg, Institute for Theoretical Physics, 
Universit\"atsstr.~31, D-93040 Regensburg, Germany}

\begin{document}

\maketitle

\begin{abstract}
The realization of center and chiral symmetries in $\mathcal{N}=1$ super Yang-Mills theory (SYM) is investigated on a four-dimensional Euclidean lattice by means of Monte Carlo methods. At zero temperature this theory is expected to confine external fundamental charges and to have a non-vanishing gaugino condensate, which breaks the non-anomalous Z$_{2\textrm{N}_{c}}$ chiral symmetry. In previous studies at finite temperatures, the phase transitions corresponding to deconfinement and chiral restoration were observed to occur at roughly the same critical temperature for SU(2) gauge group. We find further evidences for this observation from new measurements at smaller lattice spacings using the fermion gradient flow, and we discuss the agreement of our findings with conjectures from superstring theory. The implementation of the gradient flow technique allows us also to estimate, for the first time, the condensate at zero temperatures and zero gaugino mass with Wilson fermions.
\end{abstract}

\section{Introduction}

Supersymmetric Yang-Mills theories (SYM) have been a useful laboratory for the understanding of the fundamental properties of quantum field theories. For instance, the study of supersymmetry combined with gauge symmetry has led to the discovery that duality is a feature of certain extended SYM. Besides the well-known gauge-gravity duality between the conformal $\mathcal{N}=4$ SYM and string theory on the curved-space $AdS_5 \times S^{5}$, the electro-magnetic duality has been conjectured for $\mathcal{N}=2$ SYM and $\mathcal{N}=1$ supersymmetric QCD \cite{SEI94,SEI94b,INT95}. The most interesting open challenge is the search for a dual theory of QCD that would open the possibility for the analytical understanding of confinement and strong interactions. $\mathcal{N}=1$ SYM is the supersymmetric extension of the pure gauge sector of QCD describing the strong interactions of gluons and gauginos, and it shares many interesting aspects with QCD while retaining the minimal supersymmetry. There have been several theoretical predictions that this theory should exhibit confinement, chiral symmetry breaking and mass-gap \cite{DAV99,AMA88,SHI88,MOR88}. $\mathcal{N}=1$ SYM is therefore a perfect candidate for the analytical understanding of non-pertubative phenomena of non-Abelian gauge theories, like confinement and chiral symmetry breaking. 

Lattice simulations are the natural tool to explore non-perturbatively the phase diagram of strongly interacting gauge theories, such as $\mathcal{N}=1$ SYM. The study of chiral symmetry breaking using the lattice discretization requires however a certain care. The naive discretization of the fermion action represents effectively sixteen fermion-species, the so-called doublers, instead of only one fermion. As a consequence, the fermion-boson state degeneracy and supersymmetry are broken. The Wilson term removes the doublers from the physical spectrum, at the cost of breaking chiral symmetry explicitly. The chiral condensate, the order parameter for chiral symmetry breaking, acquires a divergent additive renormalization term. The study of chiral symmetry breaking of $\mathcal{N}=1$ SYM in the Wilson formulation is therefore challenging. Recently the gradient flow has been proposed as a regularization-scheme independent smoothing technique that is able to simplify drastically the renormalization of lattice bare composite operators \cite{Luscher:2010iy, Luscher:2011bx, Luscher:2013cpa}. In particular, the flowed chiral condensate is free from the additive renormalization of Wilson fermions \cite{Luscher:2013cpa}.

In recent years the gradient flow has found a growing spectrum of applications. In particular the investigations of finite temperature QCD including the chiral transition have shown the benefits of this method \cite{Taniguchi:2017ibr, Taniguchi:2016ofw}. Furthermore, a novel approach to compute operator dimensions in conformal field theories was recently proposed in \cite{Carosso:2018bmz}, which exploits the relation between the gradient flow and renormalization group transformations. The application for supersymmetric theories has been suggested in \cite{Hieda:2017sqq,Kasai:2018koz}, where it can help to renormalize the supercurrent. In several works also a supersymmetric version of the method has been developed \cite{Kikuchi:2014rla,Kadoh:2018qwg}, which might even avoid the necessity of a multiplicative renormalization of the fermions. 

In this contribution we present an extended study of the phase diagram of $\mathcal{N}=1$ SYM with the gauge group SU(2) at zero and non-zero temperature. We measure the chiral condensate expectation value at positive gradient flow time. We show strong evidences that chiral symmetry is spontaneously broken at zero temperature by a non-vanishing expectation value of the gaugino condensate. We show that at high temperature chiral symmetry is restored and that a phase transition occurs in the massless limit. We also show that there are only two phases in the mass-less limit, characterized by both chiral symmetry breaking and confinement at low temperature and chiral symmetry restoration and deconfinement at high temperature. We have not found any evidences of mixed phases where deconfinement occurs while chiral symmetry is broken. A possible interpretation of our results from the point of view of superstring/M-theory is presented in the last section.

\section{$\mathcal{N}=1$ Super Yang-Mills theory and its lattice discretization}

Supersymmetry is, according to the Coleman-Mandula theorem \cite{Coleman:1967ad}, the only possible non-trivial extension of the space-time symmetries. It extends the Poincar\'e algebra by $\mathcal{N}$ irreducible spinorial generators.
As the spinor charges act on the irreducible representations of the Poincare group by changing the spin by $\pm \frac{1}{2}$, the supermultiplets contain fermionic and bosonic fields. The simplest four-dimensional supersymmetric Yang-Mills theory corresponds to a theory with one super-multiplet, consisting on-shell and in Wess-Zumino gauge of one vector gauge field and one Majorana spinor, the so-called $\mathcal{N}=1$ super Yang-Mills theory. In the present study we focus on the gauge group SU(2). For the Euclidean metric, the four-dimensional on-shell action can be written as 
\begin{align}\label{eq:action}
S=\int d^{4}x~\left(\frac{1}{4}F^{a\mu\nu}F_{\mu\nu}^{a}+\frac{1}{2}\bar\lambda^{a}\slashed D\lambda^{a}+ m\bar\lambda^{a}\lambda^{a}+\frac{\theta}{32\pi^{2}}\epsilon_{\mu\nu\rho\sigma}F^{a\mu\nu}F^{a\rho\sigma}\right),
\end{align}
where $F$ is the usual Yang-Mills (YM) field strength, $\lambda$ a Majorana spinor, the gaugino, and $D$ the gauge covariant derivative. The mass term $m$ breaks supersymmetry softly. Here all fields are su(N$_{c}$)-algebra valued and thus transform in its adjoint representation, i.e. $a\in \{1,2,3\}$ for SU(2). The third term is a topological term, whose space-time integration yields the winding number $Q_{\textrm{top}}$ of the gauge field. An important role is played by $Q_{\textrm{top}}$ in the realization of chiral symmetry in the quantum theory, as will be discussed in the next section.

$\mathcal{N}=1$ SYM is evidently very similar to QCD, as it both lacks of scalars and includes spinor fields interacting with a gauge field. This characteristic makes the model very attractive, if one intends to get some deeper understanding of non-perturbative phenomena of QCD as chiral symmetry breaking and confinement. In supersymmetric theories these phenomena are usually more accessible due to the constrains that the supersymmetry imposes.

Observables of the theory can be numerically computed by Monte-Carlo methods and based on a discretization of the Euclidean space-time, i.e. a lattice. There, the spinors are assigned to the sites and the Yang-Mills field $A_\mu$ to the links through the parallel transporter $U_\mu=\mathrm{e}^{igaA_\mu}$ in the fundamental representation, being $a$ the lattice spacing. On the lattice, the SYM action takes on the form \cite{Veneziano:1982ah}
\begin{align*}
  S_{\mathrm{lat}}=\sum_{x}\mathrm{Re}\mathrm{tr}\left\{ \frac{\beta}{\textrm{N}_{c}}\sum_{\mu\neq\nu}\left( \frac{5}{3}P_{\mu\nu}(x)-\frac{1}{12}R_{\mu\nu}(x) \right) \right\} + \frac{1}{2}\sum_{x,y}\bar\lambda(y)D_{w}[V_{\mu}](y,x)\lambda(y)
  \end{align*}
  \begin{align*}
D_{w}(x,y)\lambda(y)=\lambda(x)-\kappa\sum_{\mu=1}^{4}\left[ (1-\gamma_{\mu})V_{\mu}(x)\lambda(x+\mu)+ (1+\gamma_{\mu})V_{\mu}^{\dagger}(x-\mu)\lambda(x-\mu)\right],
  \end{align*}
  where $V$ are the parallel transporters in the adjoint representation, defined as $V_{\mu}(x)_{ab}=2\mathrm{tr}(U_{\mu}(x)^{\dagger}\tau_{a}^{F}U_{\mu}(x)\tau_{b}^{F})$\footnote{$\tau^F$ are the generators in the fundamental representation.}. Here $\beta$ is proportional to the inverse squared gauge coupling, $P_{\mu\nu}$ is the plaquette, i.e. the minimal closed loop of parallel transporters and $R_{\mu\nu}$ is the tree-level Symanzik improvement term. One-level stout smearing with parameter $\rho=0.15$ has been employed for the links in the fermion action. Finally, $\kappa$, the hopping parameter, is equal to $\frac{1}{2m+8}$. A finite gaugino mass breaks supersymmetry softly and the bare parameter $m$ must be tuned to the point of a vanishing physical mass. A numerical accurate and inexpensive way to achieve that is by means of the adjoint-pion a-$\pi$. Although this is not a physical degree of freedom of the theory, its mass is related to the gaugino mass as $m_{\mathrm{a}-\pi}^{2}\sim m$, as was shown in \cite {MUN} within the frame of partially quenched perturbation theory.\\
  In general, the lattice discretization breaks super Poincar\'e symmetry down to some subgroup of the hypercube isometries. In the massless and continuum limit, however, $\mathcal{N}=1$ supersymmetry is recovered, as confirmed through analysis of the supersymmetry Ward identities in \cite{Bergner:2015adz}.
In thermal quantum field theory, temperature corresponds to the inverse radius of a compactificatied direction of the path integral with thermal boundary conditions, i.~e.\ periodic for bosons and antiperiodic for fermions.

\section{The phase diagram of Super Yang-Mills theory}

\subsection{Confinement of static fundamental charges}

At zero temperature, the non-supersymmetric YM vacuum is expected to behave as a confining medium for external static color-electric charges. The vacuum can be probed through the Polyakov loop in the fundamental representation of the gauge group, which is the path ordered product of the links in the fundamental representation along a line which wraps in the compact direction 
\begin{equation}
 P_L = \frac{1}{N_t V_3} \sum_{\vec{x}} \textrm{Tr}\left\{\prod_{t=1}^{N_t} U_4(\vec{x},t) \right\}\, ,
\end{equation}
where $V_3$ denotes the three-dimensional lattice volume of the non-compactified directions. 
It is not invariant under center symmetry transformations. Thus, a vanishing expectation value of the fundamental Polyakov loop implies unbroken center symmetry. 
The Polyakov loop can be related to the exponential of the free energy of an isolated static fundamental quark. Hence, a vanishing vacuum expectation value of the loop also indicates that isolated fundamental quarks are states with infinite free energy. As the temperature is increased, the vacuum expectation value of $P_L$ should become non-vanishing at some critical temperature, where center symmetry would be broken spontaneously. In the phase of broken center symmetry the energy of an isolated quark would be finite and, in that sense, one would speak of quark deconfinement.

In the case of QCD, dynamical fundamental quarks break center symmetry explicitly and there is no real confinement-deconfinement phase transition but a crossover. In the confined phase, the quark-antiquark potential grows linearly until it is screened by another quark-anti-quark pair popping up from the vacuum. In the deconfined phase, asymptotic freedom leads to a so-called quark-gluon plasma. The identification of the phase transition is much clearer for $\mathcal{N}=1$ SYM. The N-ality of the fermion representation implies that the fermion fields are singlet under center transformations. Thus, the action of the theory remains invariant with respect to center transformations even for massless gauginos and connection of confinement and center symmetry breaking is still valid. A critical behavior is expected at some finite temperature and the Polyakov loop represents a good order parameter for the deconfinement phase transition for $\mathcal{N}=1$ SYM.

\subsection{Chiral symmetry}

$\mathcal{N}=1$ supersymmetric Yang-Mills theory is invariant under certain chiral transformations when no soft SUSY-breaking mass term is included in the action. Chiral symmetry coincides with the U(1) R-symmetry of the theory, which leaves the SUSY algebra invariant\footnote{In this section $\lambda^{a}$ is a Weyl spinor, as this notation is more natural for the massless theory.}
\begin{align*}
\lambda^{a}\to\mathrm{e}^{i\alpha}\lambda^{a}\;.
\end{align*}
This symmetry is, however, broken at the quantum level by instanton contributions. In the presence of a non-vanishing $\theta$-term in the action \eqref{eq:action}, the U(1) chiral rotation is equivalent to
\begin{align*}
\theta\to \theta-2\textrm{N}_{c}\alpha \, ,
\end{align*}
where $\textrm{N}_{c}$ is the number of colors. The path integral is therefore invariant only for $\alpha=k\pi/\textrm{N}_c$, i.e.\ the chiral symmetry of the quantum theory is actually the discrete subgroup Z$_{2\textrm{N}_c}\subset\, $U(1). An interesting question is whether the quantum chiral symmetry is spontaneously broken by a non-vanishing fermion condensate. Indeed, it can be shown analytically that the answer is affirmative \cite{Witten:1982df,Veneziano:1982ah,Taylor:1982bp,Affleck:1983mk,SHI88,MOR88,AMA88,DAV99}. Thanks to remarkable properties of supersymmetric theories like non-renormalization and holomorphicity of the effective superpotential with respect to fields and couplings, it was proven \cite{Terning:2006bq, Schwetz:1997cz} that if the effective theory at long distances is massive with color-singlets as degrees of freedom, the gaugino condensate takes on the form
\begin{align*}
\langle \lambda^{a}\lambda^{a}\rangle\sim \Lambda^{3}.
\end{align*}

As the gaugino condensate is not invariant with respect to the full Z$_{2\textrm{N}_{c}}$ group, i.e. $\langle \lambda^{a}\lambda^{a}\rangle = \mathrm{e}^{2i\alpha}\langle \lambda^{a}\lambda^{a}\rangle$ only if $\alpha=0,\pi$, the vacuum of the theory does not exhibit the same symmetry as the partition function and the global Z$_{2\textrm{N}_c}$ symmetry is spontaneously broken down to its discrete subgroup Z$_{2}$, i.e. to the sign flip $\lambda^{a}\to -\lambda^{a}$. Hence, chiral symmetry is spontaneously broken at zero temperature with $\textrm{N}_{c}$ degenerate vacua, which are connected by Z$_{\textrm{N}_c}$ transformations. As the broken symmetry is a discrete one, the existence of domain walls interpolating between the different vacua is expected. In \cite{Dvali:1996xe} it was found that domain walls in SYM are BPS-saturated states and an expression of their energy density was computed.

\subsection{The phase diagram}

A non-trivial task is to determine what happens to the vacuum of the theory and specifically to the condensate and Polyakov loop distributions at finite temperature. At some critical temperature it is expected that the condensate vanishes and thus, that the vacuum becomes Z$_{2\textrm{N}_{c}}$-symmetric. In addition, center symmetry breaks spontaneously at the critical deconfinement temperature and the Polyakov loop should acquire a non-vanishing expectation value. An interesting question arising from this analysis is whether both critical temperatures coincide, i.e. if confinement and chiral symmetry breaking share the same non-perturbative origin. This question is far from trivial. Although QCD, for example, is also characterized by confinement and broken chiral symmetry at low energies, some theories have been found to exhibit only one of those \cite{Aharony:2006da}. In some numerical studies of QCD it has been found that chiral restoration and deconfinement phase transition occur at nearly the same critical temperature. However, taking QCD as a starting point implies the difficulty that these phase transitions are actually crossovers due to the finite-massive fundamental color charges. This means that no exact order parameter can be found, which unambiguously signals the pseudo-critical temperature. Hence, this observation strongly depends on the order parameters chosen. Fortunately, both the Polyakov loop and the gaugino condensate are exact order parameters for the respective phase transitions in $\mathcal{N}=1$ SYM. Consequently, the question of the relation between the chiral and deconfinement transition can be answered by studying these quantities in a systematic way, the only remaining complications being related to the regularization and the renormalization of the these observables. The lattice realization with Wilson fermions on the lattice breaks chiral symmetry, which implies besides the multiplicative renormalization factor ($Z_{\bar\lambda\lambda}$) an additional additive renormalization ($\mathbf{b_{0}}$) of the gaugino condensate
\begin{align*}
\langle\bar\lambda\lambda\rangle_{\mathrm{R}}=Z_{\bar\lambda\lambda}(\beta)(\langle \bar\lambda\lambda\rangle_{\mathrm{B}}-\mathbf{b_{0}}).
\end{align*}

In previous investigations \cite{Bergner:2014saa}, the additive constant $\mathbf{b_{0}}$ was removed by subtracting the bare condensate at zero temperature,
\begin{align*}
\langle\bar\lambda\lambda\rangle_{\mathrm{S}}=\langle\bar\lambda\lambda\rangle_{\mathrm{B}}^{T=0}-\langle\bar\lambda\lambda\rangle_{\mathrm{B}}^{T}
\end{align*}
One downside of this approach is the fact that the subtracted condensate is fixed to vanish at $T=0$, or some very low temperature. Although this subtraction should preserve the behavior of the order parameter near the critical temperature, it is not possible to determine the renormalized condensate at zero temperature. Hence, the picture of the realization of chiral symmetry in thermal SYM is, following this method, incomplete. As it will be clear in the next session, the additive renormalization is not necessary when the gradient flow is used, allowing the computation of the condensate at zero temperature and, moreover, making the results comparable to fermion discretizations which satisfy the Ginsparg-Wilson relation.

\section{The gradient flow}

\subsection{Flow equations}

Motivated in the context of trivialising maps \cite{Luscher:2009eq}, L\"uscher studied the correlation functions of fields \textit{flowed} through the equations \cite{Luscher:2011bx}  \footnote{The term $D_{\mu}\partial_{\nu}B_{\nu}$ is a gauge parameter, which is included for mere technical reasons. }
\begin{align}\label{gaugeflow}
&  \partial_{t}B_{\mu}=D_{\nu}G_{\nu\mu} + D_{\mu}\partial_{\nu}B_{\nu},\quad \left.B_{\mu}\right|_{t=0}=A_{\mu},\quad
G_{\mu\nu}=\partial_{\mu}B_{\nu}-\partial_{\nu}B_{\mu}+\left[ B_{\mu},B_{\nu} \right]\\
& \partial_{t}\chi=\Delta \chi, \quad \partial_{t}\bar{\chi}=\bar{\chi}\overleftarrow{\Delta},\quad \left.\chi\right|_{t=0}=\psi, \quad \left.\bar{\chi}\right|_{t=0}=\bar{\psi},\quad  \Delta=D_{\mu}D^{\mu}.\nonumber
\end{align}
where $B$ (resp. $A$) and $G$ are the Yang-Mills gauge field and field strength, respectively. $\chi$ and $\psi$ are spinor fields and $D_{\mu}$ is the gauge-covariant derivative in the adjoint representation. 

The parameter $t\in \mathbb{R}$ describes a flow on the vector space of gauge fields. These equations have a smoothening effect on the fields, which are Gaussian-like smeared over an effective radius $r_{t}=\sqrt{8t}$, as it can be easily seen by integrating equation (\ref{gaugeflow}) in the non-interacting limit $[B,B]\sim 0$,
\begin{align*}
B_{\mu,1}(t,x)=\int{d^{D}y~K_{t}(x-y)A_{\mu}(y)}, \quad K_{t}(z)=\int{\frac{d^{D}p}{(2\pi)^{D}}\mathrm{e}^{ipz}\mathrm{e}^{-tp^{2}}}=\frac{\mathrm{e}^{-z^{2}/4t}}{(4\pi t)^{D/2}},
\end{align*}
where $D$ is the number of space-time dimensions and $K$ the heat-kernel. Moreover, the term $\mathrm{e}^{-tp^{2}}$ regularizes the integral in momentum space when $t>0$, removing the UV divergences at large momenta. Some years ago it was shown that through this kind of gradient flow the smearing property remains at all orders in perturbation theory \cite{Luscher:2011bx}. In this context, L\"uscher developed a  D+1 dimensional quantum field theory, where $t$ is taken as an spurious extra Euclidean dimension.\footnote{ The unflowed theory lives on the D-dimensional boundary of a D+1 dimensional manifold, where fields propagate through the kernel of the flow equations.} Fields would propagate along the new dimension through the heat kernel, which is a retarded propagator (there are no \textit{flow loops}) and the flow equations are imposed through Lagrange-multiplier-fields. It was found that in this theory the BRS-Ward identities hold only if there are no counter-terms beyond those of the unflowed theory.  This led to the observation that correlation functions of monomials of flowed gauge-fields are renormalized without the necessity of extra counter-terms and that spinor operators renormalize multiplicatively \cite{Luscher:2013cpa}. \footnote{The monomials renormalize according to the field content since the flowed fields are effectively non-local.} Furthermore, a specific advantage of the gradient flow relies on the fact that it is regularization-scheme independent and all these results are expected to hold also on a space-time lattice. Therefore, currents and densities which are explicitly broken by the lattice discretization should be more easily accessible within this method. As a special case, the additive renormalization constant, necessary for the computation of the gaugino condensate with Wilson fermions, is not required if these are flowed up to some finite flow-value. Consequently, it is possible to study if $\langle \bar\lambda\lambda \rangle\neq 0$ at $T=0$ even with Wilson fermions. The value of $Z_{\bar\lambda\lambda}$ is irrelevant if the bare lattice gauge coupling is fixed. Since usual lattices are Euclidean, non-zero temperatures are achieved just by compactifying and changing the number of lattice sites in temporal direction, where fermion fields fulfill thermal (anti-periodic) boundary conditions.

\subsection{Computation of the chiral condensate from the gradient flow}

The flowed bare gaugino condensate can be defined as \cite{Luscher:2013cpa}
\begin{align}\label{contcond}
\langle \bar\lambda(x,t)\lambda(x,t)\rangle_{\mathrm{B}}=-\int{d^{D}v d^{D}w~K(t,x;0,v)S(v,w)K(t,x;0,w)^{\dagger}}.
\end{align}
Hence, computing the flowed condensate amounts to the action of the heat-kernel on the fermion propagator $S$. Analogously, its discrete version can be straightforwardly written:
\begin{align*}
\langle \bar\lambda(x,t)\lambda(x,t)\rangle_{\mathrm{B}}=-\sum_{v,w}\langle{\mathrm{tr}\{K(t,x;0,v)(D_{\mathrm{W}}(v,w))^{-1}K(t,x;0,w)^{\dagger}\}\rangle},
\end{align*}

where $D_{\mathrm{W}}$ is the Wilson-Dirac operator and the trace runs over space-time (spinor) and gauge group indices. Following \cite{Luscher:2013cpa}, the trace is estimated stochastically by inserting a complete set of random complex vectors $\eta(x)$ with $\langle \eta(x) \rangle=0$ and $\langle \eta(x)\eta(y)^{\dagger}\rangle=\delta_{x,y}$:

\begin{align*}
 & \langle\bar\lambda\lambda(t) \rangle=\frac{1}{N_{\Gamma}}\sum_{x\in \Gamma}{\langle \bar\lambda\lambda(x,t)}\rangle=-\frac{1}{N_{\Gamma}}\sum_{v,w}\langle \xi(t;0,v)^{\dagger}D^{-1}_{\mathrm{W}}\xi(t;0,w)\rangle,\\
  &\xi(t;s,w)=\sum_{x}K(t,x;s,w)^{\dagger}\eta(x).
\end{align*}

Here $\langle \dots \rangle$ denotes the average with respect to to both the Monte-Carlo time and any internal group-symmetry representations. Finally, to compute the new vectors $\xi$, the so-called \textit{adjoint} flow equation 

\begin{align}
\label{adjfloweq}
\partial_{s}\xi(t;s,w)=-\Delta(V_{s})\xi(t;s,w), \quad \xi(t;t,w)=\eta(w),
\end{align}

must be integrated from $s=t$ to $s=0$, i.e. backwards in comparison with the flow equations presented above. Here, the gauge connection $V_{s}$ in the covariant four-dimensional Laplacian $\Delta$ is flowed up to the same $t$ as $\eta$.\\

The proposed path for the numerical computation is as follows \cite{Luscher:2013cpa,Taniguchi:2016ofw}. First, one initial gauge configuration $V_{t=0}$ is picked and flowed by integrating the discrete version of the gauge part of Eq.(\ref{gaugeflow})

\begin{align*}
\dot{V}_{t}(x,\mu)=-\left\{ \partial_{x,\mu}S_{\mathrm{w}}(V_{t}) \right\}V_{t}(x,\mu),\quad \left.V_{t}(x,\mu)\right|_{t=0}=U(x,\mu).
\end{align*}

up to some $t$ by means of the following Runge-Kutta integrator with step-size $\epsilon$

\begin{align*}
  W_{0}&=V_{t},\\
  W_{1}&=\exp{(1/4)Z_{0}}W_{0},\\
  W_{2}&=\exp{(8/9)Z_{1}- (17/36)Z_{0}}W_{1},\\
  V_{t+\epsilon}&=\exp{(3/4)Z_{2}-(8/9)Z_{1}+(17/36)Z_{0}}W_{2},~\mathrm{with}~~Z_{i}=-\epsilon \partial_{x,\mu}S_{\mathrm{w}}(W_{i}),
\end{align*}

where S$_{w}$ is the Wilson plaquette action. The intermediate fields are then kept on the computer memory in order to access them during the integration of the adjoint fermion equation.  As a second step one random source vector $\xi_{s=t}=\eta$ is generated on the lattice and then integrated by means of the Runge-Kutta integrator down to $s=0$

\begin{align*}
  \lambda_{3}&=\xi^{\epsilon}_{s+\epsilon},\\
  \lambda_{2}&=\frac{3}{4}\Delta_{2}\lambda_{3},\\
  \lambda_{1}&=\lambda_{3} + \frac{8}{9}\Delta_{1}\lambda_{2},\\
  \lambda_{0}&=\lambda_{1} + \lambda_{2} + \frac{1}{4}\Delta_{0}\left( \lambda_{1}-\frac{8}{9}\lambda_{2} \right),~~~ \mathrm{with}~~~\xi^{\epsilon}_{s}=\lambda_{0}
\end{align*}

Further, the vector $W(v)=\sum_{w}D(v,w)^{-1}\xi(t;0,w)$ is computed, e.g. through conjugate gradient. Subsequently it is contracted with $\xi(t;0,v)^{\dagger}$ and the result averaged over the lattice sites, i.e. one calculates $ -\frac{1}{N_{lat}}\sum_{v}{\xi_{k}(t;0,v)^{\dagger}W(v)}$.

\section{Results}

\subsection{Gaugino condensate at zero temperature}
\begin{figure}[htbp!]
  \centering
  \hspace*{-1.3cm}
  \captionsetup{font=footnotesize} 
  \scalebox{0.9}{\input{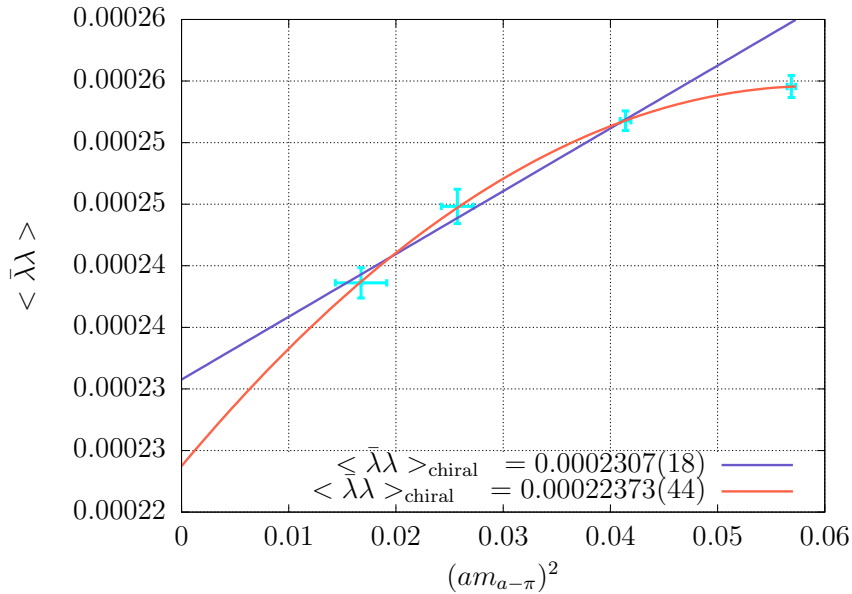}}
  \caption{Extrapolation to vanishing renormalised gaugino mass. It can be seen that the gaugino condensate scales almost linearly with the lattice spacing $a$. Some weak quadratic dependence is noticeable as the gaugino mass grows. Most important, the condensate does not vanish at the chiral point, which implies the spontaneously symmetry breaking of the Z$_{2\textrm{N}_{c}}$ symmetry.}
  \label{chiext}
  \end{figure}
As anticipated, the remarkable properties of the gradient flow method allow for the computation of the gaugino condensate at zero temperature, since no ambiguity is introduced by the additive renormalization. We have taken for ensembles from our previous investigations of SU(2) SYM at $\beta=1.75$ on a $32^{3}\times 64$ lattice with Hopping-parameter values \[\kappa \in \{0.1490,0.1492,0.1494,0.1495\}.\] The data of these four points were then used to make an extrapolation to the chiral point, i.e. to vanishing renormalized gaugino mass. As shown in Fig. \ref{chiext}, we obtain a finite value of the condensate in the chiral limit. 
For the numerical simulations with Wilson fermions, this is the first clear indication that the discrete chiral symmetry is spontaneously broken at zero temperature and zero gaugino mass. 
In previous studies without the method of the gradient flow, only an indirect observation of the non-zero condensate has been possible from a double peak of the histogram that appears in rather unstable simulations close to the chiral point \cite{Kirchner:1998mp,Ali:2018dnd}. 
Our result is remarkably compatible with the results obtained through Domain-Wall and overlap fermions in \cite{Fleming:2000fa,Giedt:2008xm,Endres:2009yp,Giedt:2009yd,Kim:2011fw}. However, for a direct comparison of the results a consistent renormalization of the condensate obtained with the Domain-Wall fermions would be required. Although the existence of a non-vanishing condensate is a well-known fact, being able to compute it from Wilson fermions without the complication of a residual mass, shows the goodness of the gradient flow method. Moreover it opens up the possibility to study the different phases at zero temperature without the computational more expensive Ginsparg-Wilson fermions.
\subsection{Discrete chiral symmetry restoration and quark deconfinement}

The second main purpose of this study is to investigate the realization of the chiral and center symmetries in SU(2) SYM as temperature is turned on. Some first results in this direction were obtained in \cite{Bergner:2014saa}. This problem now revisited using the gradient flow on four different new lattice ensembles at $\beta=1.75$ and Hopping parameter $\kappa\in \{0.1480,0.1490,0.14925\}$. To further cross-check the validity of our results, we have also analyzed a set of ensembles generated at $\beta=1.65$ and $\kappa=0.175$ using the tree-level clover improved fermion action with unsmeared links. Finite temperatures were achieved by fixing the lattice parameters and compactifying one dimension on a circle, imposing thermal boundary conditions on the fields, i.e. anti(periodic) for fermions (bosons). Thus, the lattice size was set to $24^{3}\times N_{t}$ for $N_{t}\in \{4,5,6,\cdots,48\}$, with the upper bound parametrizing the zero-temperature limit. Setting the scale, i.e. going from lattice to physical units, was achieved through the $t_{0}$ parameter, which is defined through the gradient flow as \footnote{Here a mass-dependent scheme was chosen, i.e. a different $t_{0}$ for each $\kappa$ value.}

\begin{align*}
  \left.t^{2}\langle \varepsilon(t)\rangle\right|_{t=t_{0}}=0.3, \quad \varepsilon=\frac{1}{4}G_{\mu\nu}^{a}G_{\mu\nu}^{a},
\end{align*}

with $\varepsilon$ the field energy density and $G_{\mu\nu}^{a}$ the flowed Yang-Mills field strength.\\

\begin{figure}[h!]
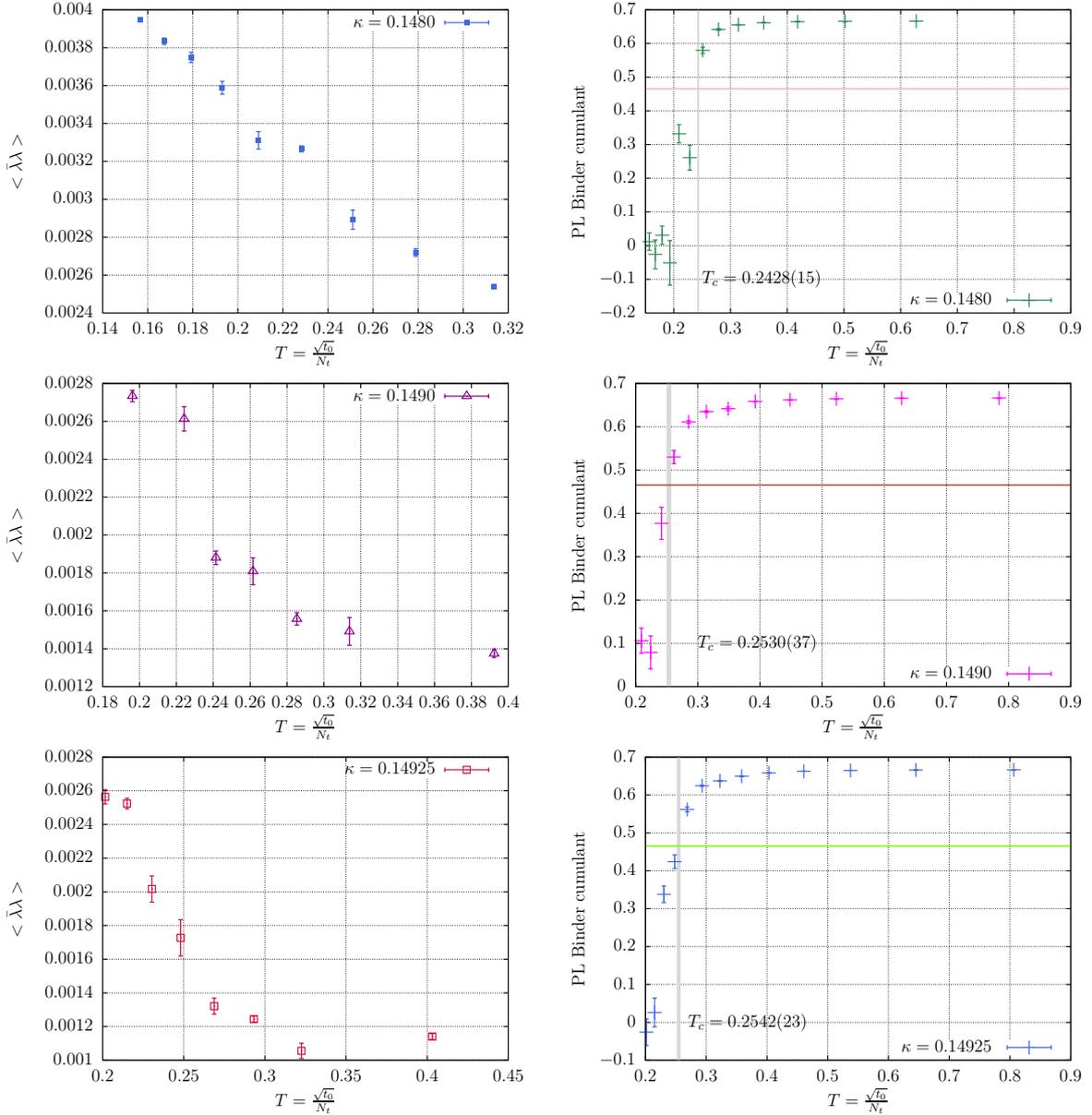

  \centering
  \captionsetup{font=footnotesize} 
  \begin{subfigure}[b]{0.5\linewidth}
      \scalebox{.6}{\input{./figures/condensate-t0.tex}}
  \end{subfigure}
  \begin{subfigure}[b]{0.49\linewidth}
    \scalebox{.6}{\input{./figures/PL-175b_1480k.tex}}
  \end{subfigure}

  \begin{subfigure}[b]{0.5\linewidth}
    \scalebox{.6}{\input{./figures/condensate1490.tex}}
  \end{subfigure}
  \begin{subfigure}[b]{0.49\linewidth}
    \scalebox{.6}{\input{./figures/PL-175b_1490k.tex}}
  \end{subfigure}

  \begin{subfigure}[b]{0.5\linewidth}
    \scalebox{.6}{\input{./figures/condensate14925.tex}}
  \end{subfigure}
  \begin{subfigure}[b]{0.49\linewidth}
    \scalebox{.6}{\input{./figures/PL-175b_14925k.tex}}
  \end{subfigure}
  \caption{On the left side the temperature dependence of the bare gaugino condensate is shown for $\kappa=0.1480,0.1490$ and $0.14925$, with the smallest $\kappa$ in the uppermost plot. On the undermost graphic, the point corresponding to the highest temperature appears to show a growth in the condensate. This is however a non-physical over-smoothing artefact due to the fact that $\sqrt{8t_{0}}> N_{t}$ in that region. The right-hand side shows the Binder cumulant of the Polyakov loop for the same $\kappa$ values.   }
  \label{su2}
\end{figure}

\begin{figure}
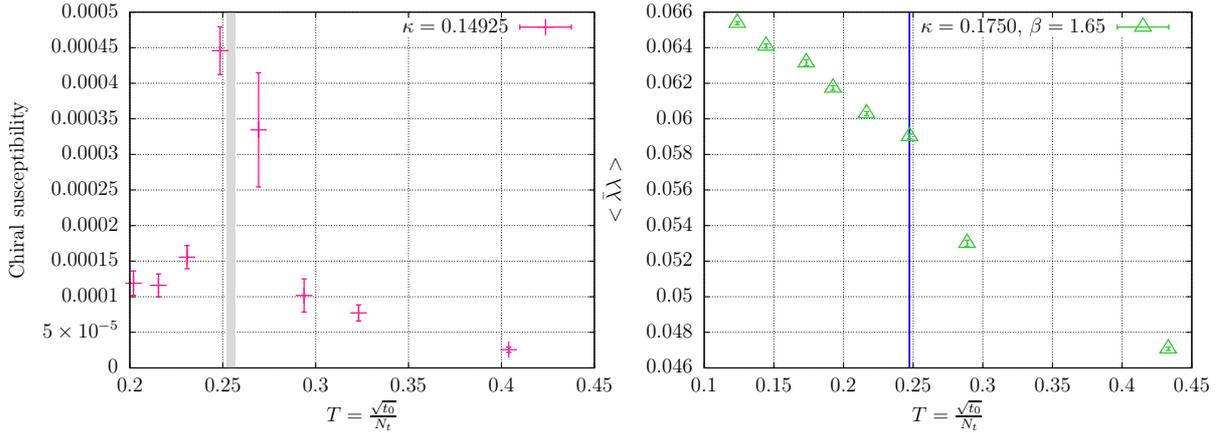

  \centering
    \captionsetup{font=footnotesize} 
  \begin{subfigure}[b]{0.49\linewidth}
  \scalebox{0.65}{\input{./figures/susc-14925k.tex}}
\end{subfigure}
\begin{subfigure}[b]{0.49\linewidth}
  \scalebox{0.65}{\input{./figures/beta165kappa0175.tex}}
  \end{subfigure}
  \caption{On the left: Chiral susceptibility for $\kappa=0.14925$. The gray band denotes the critical temperature of the Polyakov loop. On the right: gaugino condensate at $\beta=1.65$ and $\kappa=0.175$. The phase transition occurs at $N_{t}^c=7$, which roughly agrees in dimensionless units with the critical temperature for $\beta=1.75$.}
    \label{susc}
\end{figure}

The results are summarized in figure \ref{su2} and \ref{susc}. The condensate is in general considerably reduced as the temperature increases, but a clear phase transition is difficult to identify for the larger gaugino masses corresponding to smaller values of $\kappa$. This smooth behavior of the chiral condensate at larger masses is expected since the jump of the order parameter around the pseudo-critical temperature is less pronounced for a larger explicit symmetry breaking. For the smallest gaugino mass ($\kappa=0.14925$) the signal is considerably better and a jump at the critical temperature can be identified. This suggests, as expected, that the chiral restoration becomes indeed a true phase transition and the gaugino condensate an adequate order parameter in the chiral (and supersymmetric) limit. This can be more clearly seen from the disconnected chiral susceptibility in Figure \ref{susc}.\footnote{The disconnected part of the susceptibility is expected to represent the largest contribution to the phase transition's peak \cite{Bergner:2014saa}} It is remarkable that the phase transition appears to happen at $T=\sqrt{t_{0}}/N_{t}\sim 0.25$ also with the improved lattice action, 
see Figure \ref{susc}.

This temperature approximately coincides with the deconfinement phase transition, which was found in \cite{Bergner:2014saa} to be second order and to have the critical behavior of the Z$_{2}$ Ising model. The deconfinement transition occurs thus at the point where the Binder cumulant of the Polyakov loop 
\begin{align*}
B_{4}(P_{L})=1-\frac{1}{3}\frac{\langle P_L^{4} \rangle}{\langle P_L^{2}\rangle^{2}}
\end{align*}
reaches the critical value 0.46548(5) \cite{Ferrenberg:2018zst}. This point is shown in the plots on the right side of Figure \ref{su2}. As the vanishing of the condensate in the chiral and continuum limit also signals a true phase transition at this point, our results show that, up to numerical uncertainties, in $\mathcal{N}=1$  SU(2) supersymmetric Yang-Mills theory, the deconfinement phase transition and restoration of the discrete Z$_{2\textrm{N}_{c}}$ chiral symmetry occur simultaneously. This fact is far from trivial. Indeed, some non-supersymmetric QCD-like theories, like YM with several adjoint spinors, have been observed to exhibit phases with mixed deconfining and broken chiral symmetry phases \cite{Karsch:1998qj}.\\
Our results are furthermore in agreement with the predictions from Ref \cite{Gaiotto:2017yup,Komargodski:2017smk}. There the authors show, through anomaly matching arguments, that unbroken center symmetry must imply broken Z$_{2\mathrm{N}_{c}}$ chiral symmetry in $\mathcal{N}=1$ supersymmetric Yang-Mills, meaning that $T_{\mathrm{chiral}}^{c}\geq T_{\mathrm{deconfinement}}^{c}$. The same constraint was also found from anomaly matching for SU(N$_{c}$) with adjoint and fundamental matter in \cite{Shimizu:2017asf}. Our results indicate that this inequality is saturated, as predicted remarkably for adjoint QCD and SYM on $\mathbb{R}^{3}\times S^{1}$ in \cite{Anber:2011gn,Anber:2012ig,Anber:2013doa,Anber:2018ohz}. 

An interesting final quantity determined by our study is the deconfinement temperature in the chiral/supersymmetric limit. We compare our result to the deconfinement critical temperature of SU(2) YM found in \cite{Cella:1993ic} through the ratio 
\begin{align*}
\frac{T_{c}(\mathrm{SYM})}{T_{c}(\mathrm{YM})}=\frac{0.2574(3)}{0.3082(2)}=0.8354(12)\,,
\end{align*}
where the two temperatures are compared in dimensionless units as $\sqrt{t_0}/N_t^c$, and the scale $t_0$ for pure gauge has been computed in \cite{GP}. This value is compatible with our previous investigations in \cite{Bergner:2014saa} at a smaller $\beta$. The ratio furthermore roughly agrees with an analytical prediction found in \cite{Lacroix:2014jpa}, where it is claimed that $\frac{T_{c}(\mathrm{SYM})}{T_{c}(\mathrm{YM})}=\sqrt{\frac{2}{3}}\sim 0.8$.

\subsection{Prediction from string theory}

The result that deconfinement and restoration of the quantum chiral symmetry are occurring at the same critical point is far from trivial and can only be confirmed through non-perturbative methods. Lattice simulations can compute an estimation of the order parameters, but they do not provide immediately a qualitative physical interpretation of the mechanisms responsible for such results. A deep understanding of the dynamics of non-Abelian gauge theories in general and QCD in particular is, indeed, still missing and concepts like supersymmetry and string theory arose in the attempt to achieve it. 
An interpretation developed in the context of string theory might indeed provide such a deeper understanding of the mechanism behind our results. 

In \cite{Witten:1997ep} Witten considered certain brane configuration consisting of two differently oriented NS five-branes and $\textrm{N}_{c}$ D-four-branes stretching between them in weakly coupled IIA superstring theory.\footnote{The framework is actually M-theory. The brane model is not equivalent to SYM but is in the same universality class. SYM is obtained when taking the IIA, i.e.\ ten dimensional limit.} The effective theory on the world-volume of the four-brane is a 3+1 dimensional SU(N$_{c}$) gauge theory with $\mathcal{N}=1$ supersymmetry. This theory has $\textrm{N}_{c}$ vacua, which can be identified with the pattern of chiral symmetry breaking of the supersymmetric QFT. Witten succeeded to show that the confining string emanating from an external quark is topologically equivalent to the IIA fundamental string. Furthermore he found that the BPS-saturated domain wall can actually be identified with a D-brane, on which the confining strings can end. This fact directly relates confinement and chiral symmetry breaking. Indeed, the domain wall is expected to have restored chiral symmetry in its core, while a color-electric source sufficiently near the wall behaves as a free quark, and the Polyakov loop expectation value does not vanish. This was investigated further in \cite{Campos:1998db} by means of an effective field theory for the gaugino condensate and the Polyakov loop. The authors found that, for $\textrm{N}_{c}=3$, Z$_{3}^{\chi}$ restoration implies Z$_{3}^{c}$ breaking and that the results of Witten, i.e. that the confining strings end on the domain wall, can only hold if both phase transitions occur simultaneously.

\section{Conclusions and Outlook}

The gradient flow has enabled us to explore the confining and the chiral properties of the $\mathcal{N}=1$ SYM both at zero and finite temperature. We have been able to extrapolate the chiral condensate at zero temperature to the mass-less limit and to prove that chiral symmetry is broken. Our findings with the Wilson fermion action are in agreement with previous studies with Ginsparg-Wilson fermions, avoiding, however, the huge numerical cost of preserving chiral symmetry on the lattice. A precise quantitative comparison of the results obtained with the different fermion actions would be possible once a common renormalization scheme would be chosen to fix the multiplicative renormalization constant. 

We have also explored the phase diagram of the theory at non-zero temperature. The chiral condensate develops an abrupt change of behavior as the gaugino mass becomes smaller, compatible with a second order phase transition. In comparison to our previous investigations without gradient flow in \cite{Bergner:2014saa}, the jump in the order parameter's expectation value is clearer and thus more significant, which reflects the advantages of the method. Further, the study of the Binder cumulant of the Polyakov loop has been important to locate precisely the deconfinement phase transition. Through the identification of both phase transitions we found that the critical temperatures are in fact coincident. This leads us to the remarkable conclusion that chiral symmetry restoration and deconfinement are not independent and uncorrelated non-perturbative phenomena in $\mathcal{N}=1$ supersymmetric Yang-Mills theory, but they may obey a common underlying dynamics. As pointed out in the end of the present work, this observation seems to be in agreement with previous semiclassical predictions found through certain brane configuration in IIA superstring/M-theory.

We are currently working towards new applications of the gradient flow. One immediate further step is to investigate thermal SYM for SU(3) gauge group. Another very interesting direction is to take profit of the method to study the vacuum's structure of the theory at zero temperature. In addition we plan to investigate possible applications for the renormalization of the supercurrent in theories with extended supersymmetry \cite{Kasai:2018koz}.

\section{Acknowledgements}
The configurations and the parameters of this study have been created in the long term effort of the DESY-Münster collaboration to study the properties of
SYM. We thank in particular Gernot Münster and Istvan Montvay for helpful comments and discussions.
The authors gratefully acknowledge the Gauss Centre for Supercomputing
e.~V.\,\linebreak(www.gauss-centre.eu) for funding this project by providing
computing time on the GCS Supercomputers JUQUEEN, JURECA, and JUWELS at J\"ulich Supercomputing
Centre (JSC) and SuperMUC at Leibniz Supercomputing Centre (LRZ).
G.~Bergner and C.~L\'opez acknowledge support from the Deutsche
Forschungsgemeinschaft (DFG) Grant No.\ BE 5942/2-1.

\pagebreak


\begin{thebibliography}{100}

\bibitem{SEI94}
N. Seiberg and E. Witten, \emph{Electric - magnetic duality, monopole condensation, and confinement in N=2 supersymmetric Yang-Mills theory}, Nucl. Phys. B426 (1994) 19-52, \arxivold{hep-th/9407087}.

\bibitem{SEI94b}
N. Seiberg, \emph{Electric-Magnetic Duality in Supersymmetric Non-Abelian Gauge Theories}, Nucl. Phys. B435(1995) 129-146, \arxivold{hep-th/9411149}.

\bibitem{INT95}
K. Intriligator, N. Seiberg, \emph{Lectures on supersymmetric gauge theories and electric-magnetic duality}, Nucl. Phys. Proc. Suppl. 45BC:1-28 (1996), \arxivold{hep-th/9509066}.




\bibitem{DAV99}
N. M. Davies, T. J. Hollowood, V. V. Khoze, M. P. Mattis:
\emph{Gluino Condensate and Magnetic Monopoles in Supersymmetric Gluodynamics},
Nucl.\ Phys.\ B {\bf 559} (1999) 123,
\arxivold{hep-th/9905015}.

\bibitem{AMA88}
D.~Amati, K.~Konishi, Y.~Meurice, G.~C.~Rossi and G.~Veneziano:
\emph{Nonperturbative Aspects in Supersymmetric Gauge Theories},
Phys.\ Rept.\ {\bf 162} (1988) 169.


\bibitem{SHI88}
M. A. Shifman and A. I. Vainshtein,
\emph{On Gluino Condensation in Supersymmetric Gauge Theories. SU(N) and O(N) Groups},
Nucl.\ Phys.\ B {\bf 296} (1988) 445.

\bibitem{MOR88}
A. Y. Morozov, M. A. Olshanetsky and M. A. Shifman,
\emph{Gluino condensate in supersymmetric gluodynamics},
Nucl.\ Phys.\ B {\bf 304} (1988) 291.

\bibitem{Luscher:2010iy}
  M.~L\"uscher,
  \emph{Properties and uses of the Wilson flow in lattice QCD},
  JHEP {\bf 1008} (2010) 071
   Erratum: [JHEP {\bf 1403} (2014) 092]
   \arxiv{1006.4518}{\tt hep-lat}.

\bibitem{Luscher:2011bx}
  M.~L\"uscher and P.~Weisz,
\emph{Perturbative analysis of the gradient flow in non-abelian gauge theories},
  JHEP {\bf 1102} (2011) 051
  \arxiv{1101.0963}{hep-th}.

\bibitem{Luscher:2013cpa}
  M.~L\"uscher,
\emph{Chiral symmetry and the Yang--Mills gradient flow},
  JHEP {\bf 1304} (2013) 123
  \arxiv{1302.5246}{hep-lat}.

  \bibitem{Taniguchi:2017ibr}
Y.~Taniguchi {\it et al.} [WHOT-QCD Collaboration],
\emph{Energy-momentum tensor correlation function in $N_f = 2 + 1$ full QCD at finite temperature},
\arxiv{1711.02262}{hep-lat}.

\bibitem{Taniguchi:2016ofw}
Y.~Taniguchi, S.~Ejiri, R.~Iwami, K.~Kanaya, M.~Kitazawa, H.~Suzuki, T.~Umeda and N.~Wakabayashi,
\emph{Exploring $N_{f}$ = 2+1 QCD thermodynamics from the gradient flow},
Phys.\ Rev.\ D {\bf 96} (2017) no.1,  014509


\bibitem{Carosso:2018bmz}
A.~Carosso, A.~Hasenfratz and E.~T.~Neil,
\emph{Nonperturbative Renormalization of Operators in Near-Conformal Systems Using Gradient Flows},
Phys.\ Rev.\ Lett.\  {\bf 121} (2018) no.20,  201601
\arxiv{1806.01385}{hep-lat}.

\bibitem{Hieda:2017sqq}
  K.~Hieda, A.~Kasai, H.~Makino and H.~Suzuki,
  \emph{4D $\mathcal{N}=1$ SYM supercurrent in terms of the gradient flow},
  PTEP {\bf 2017} (2017) no.6,  063B03
  \arxiv{1703.04802}{hep-lat}.
  
\bibitem{Kasai:2018koz}
A.~Kasai, O.~Morikawa and H.~Suzuki,
\emph{Gradient flow representation of the four-dimensional $\mathcal{N}=2$ super Yang–Mills supercurrent},
PTEP {\bf 2018} (2018) no.11,  113B02
\arxiv{1808.07300}{ hep-lat}.

\bibitem{Kikuchi:2014rla}
  K.~Kikuchi and T.~Onogi,
  \emph{Generalized Gradient Flow Equation and Its Application to Super Yang-Mills Theory},
  JHEP {\bf 1411} (2014) 094
  \arxiv{1408.2185}{hep-th}.

\bibitem{Kadoh:2018qwg}
D.~Kadoh and N.~Ukita,
\emph{Supersymmetric gradient flow in $\mathcal N=1$ SYM},
\arxiv{1812.02351}{hep-th}.

\bibitem{Coleman:1967ad}
S.~R.~Coleman and J.~Mandula,
\emph{All Possible Symmetries of the S Matrix},
Phys.\ Rev.\  {\bf 159} (1967) 1251.

\bibitem{Veneziano:1982ah}
  G.~Veneziano and S.~Yankielowicz,
  \emph{An Effective Lagrangian for the Pure N=1 Supersymmetric Yang-Mills Theory},
  Phys.\ Lett.\  {\bf 113B} (1982) 231.

\bibitem{MUN}
  G. M\"unster, H. St\"uwe, \emph{The mass of the adjoint pion in N=1 supersymmetric Yang-Mills theory}, JHEP05(2014)034, \arxiv{1402.6616}{hep-th}.

\bibitem{Bergner:2015adz}
G.~Bergner, P.~Giudice, G.~Münster, I.~Montvay and S.~Piemonte,
\emph{The light bound states of supersymmetric SU(2) Yang-Mills theory},
JHEP {\bf 1603} (2016) 080  
  
\bibitem{Witten:1982df}
  E.~Witten,
  \emph{Constraints on Supersymmetry Breaking},
  Nucl.\ Phys.\ B {\bf 202} (1982) 253.


\bibitem{Taylor:1982bp}
  T.~R.~Taylor, G.~Veneziano and S.~Yankielowicz,
  \emph{Supersymmetric QCD and Its Massless Limit: An Effective Lagrangian Analysis},
  Nucl.\ Phys.\ B {\bf 218} (1983) 493.

  
\bibitem{Affleck:1983mk}
  I.~Affleck, M.~Dine and N.~Seiberg,
  \emph{Dynamical Supersymmetry Breaking in Supersymmetric QCD},
  Nucl.\ Phys.\ B {\bf 241} (1984) 493.


\bibitem{Terning:2006bq}
  J.~Terning,
  \emph{Modern supersymmetry: Dynamics and duality},
  International series of monographs on physics. 132

\bibitem{Schwetz:1997cz}
M.~Schwetz and M.~Zabzine,
\emph{Gaugino condensate and Veneziano-Yankielowicz effective Lagrangian},
\arxivold{hep-th/9710125}


\bibitem{Dvali:1996xe}
G.~R.~Dvali and M.~A.~Shifman,
\emph{Domain walls in strongly coupled theories},
Phys.\ Lett.\ B {\bf 396} (1997) 64
Erratum: [Phys.\ Lett.\ B {\bf 407} (1997) 452]
 


\bibitem{Aharony:2006da}
O.~Aharony, J.~Sonnenschein and S.~Yankielowicz,
\emph{A Holographic model of deconfinement and chiral symmetry restoration}
Annals Phys.\  {\bf 322} (2007) 1420
\arxivold{hep-th/0604161}.

\bibitem{Bergner:2014saa}
G.~Bergner, P.~Giudice, G.~Münster, S.~Piemonte and D.~Sandbrink,
\emph{Phase structure of the $ \mathcal{N}=1 $ supersymmetric Yang-Mills theory at finite temperature},
JHEP {\bf 1411} (2014) 049

\bibitem{Luscher:2009eq}
M.~L\"uscher,
\emph{Trivializing maps, the Wilson flow and the HMC algorithm},
Commun.\ Math.\ Phys.\  {\bf 293} (2010) 899
  \arxiv{0907.5491}{hep-lat}.

\bibitem{Kirchner:1998mp}
  R.~Kirchner {\it et al.} [DESY-Munster Collaboration],
  \emph{Evidence for discrete chiral symmetry breaking in N=1 supersymmetric Yang-Mills theory},
  Phys.\ Lett.\ B {\bf 446} (1999) 209
  \arxivold{hep-lat/9810062}.

\bibitem{Ali:2018dnd}
  S.~Ali, G.~Bergner, H.~Gerber, P.~Giudice, I.~Montvay, G.~Münster, S.~Piemonte and P.~Scior,
  \emph{The light bound states of $\mathcal{N}=1$ supersymmetric SU(3) Yang-Mills theory on the lattice},
  JHEP {\bf 1803} (2018) 113
  \arxiv{1801.08062}{hep-lat}.


\bibitem{Fleming:2000fa}
  G.~T.~Fleming, J.~B.~Kogut and P.~M.~Vranas,
  \emph{SuperYang-Mills on the lattice with domain wall fermions},
  Phys.\ Rev.\ D {\bf 64} (2001) 034510
  doi:10.1103/PhysRevD.64.034510
  \arxivold{hep-lat/0008009}.

  
\bibitem{Giedt:2008xm}
  J.~Giedt, R.~Brower, S.~Catterall, G.~T.~Fleming and P.~Vranas,
  \emph{Lattice super-Yang-Mills using domain wall fermions in the chiral limit},
  Phys.\ Rev.\ D {\bf 79} (2009) 025015
  \arxiv{0810.5746}{hep-lat}.

\bibitem{Endres:2009yp}
  M.~G.~Endres,
  \emph{Dynamical simulation of N=1 supersymmetric Yang-Mills theory with domain wall fermions},
  Phys.\ Rev.\ D {\bf 79} (2009) 094503
  \arxiv{0902.4267}{hep-lat}.
  
\bibitem{Giedt:2009yd}
J.~Giedt,
\emph{Progress in four-dimensional lattice supersymmetry},
Int.\ J.\ Mod.\ Phys.\ A {\bf 24} (2009) 4045
\arxiv{0903.2443}{hep-lat}.

  
\bibitem{Kim:2011fw}
  S.~W.~Kim {\it et al.} [JLQCD Collaboration],
  \emph{Lattice study of 4d {\cal N}=1 super Yang-Mills theory with dynamical overlap gluino},
  PoS LATTICE {\bf 2011} (2011) 069
  \arxiv{1111.2180}{hep-lat}.

\bibitem{Ferrenberg:2018zst}
A.~M.~Ferrenberg, J.~Xu and D.~P.~Landau,
\emph{Pushing the limits of Monte Carlo simulations for the three-dimensional Ising model},
Phys.\ Rev.\ E {\bf 97} (2018) no.4,  043301
\arxiv{1806.03558}{physics.comp-ph}.  

\bibitem{Karsch:1998qj}
F.~Karsch and M.~Lutgemeier,
\emph{Deconfinement and chiral symmetry restoration in an SU(3) gauge theory with adjoint fermions},
Nucl.\ Phys.\ B {\bf 550} (1999) 449

\bibitem{Gaiotto:2017yup}
D.~Gaiotto, A.~Kapustin, Z.~Komargodski and N.~Seiberg,
\emph{Theta, Time Reversal, and Temperature},
JHEP {\bf 1705} (2017) 091
\arxiv{1703.00501}{hep-th}.

\bibitem{Komargodski:2017smk}
Z.~Komargodski, T.~Sulejmanpasic and M.~Ünsal,
\emph{Walls, anomalies, and deconfinement in quantum antiferromagnets},
Phys.\ Rev.\ B {\bf 97} (2018) no.5,  054418
\arxiv{1706.05731}{cond-mat.str-el}.

\bibitem{Shimizu:2017asf}
H.~Shimizu and K.~Yonekura,
\emph{Anomaly constraints on deconfinement and chiral phase transition},
Phys.\ Rev.\ D {\bf 97} (2018) no.10,  105011
\arxiv{1706.06104}{hep-th}.

\bibitem{Anber:2011gn}
M.~M.~Anber, E.~Poppitz and M.~Unsal,
\emph{2d affine XY-spin model/4d gauge theory duality and deconfinement},
 JHEP {\bf 1204} (2012) 040
\arxiv{1112.6389}{hep-th}.
  
\bibitem{Anber:2012ig}
M.~M.~Anber, S.~Collier and E.~Poppitz,
\emph{The $SU(3)/Z_3$ QCD(adj) deconfinement transition via the gauge theory/'affine' XY-model duality},
JHEP {\bf 1301} (2013) 126
\arxiv{1211.2824}{hep-th}.

\bibitem{Anber:2013doa}
M.~M.~Anber, S.~Collier, E.~Poppitz, S.~Strimas-Mackey and B.~Teeple,
\emph{Deconfinement in $\mathcal{N}=1$ super Yang-Mills theory on $\mathbb{R}^3 \times \mathbb{S}^1$ via dual-Coulomb gas and "affine" XY-model},
JHEP {\bf 1311} (2013) 142
\arxiv{1310.3522}{hep-th}.

\bibitem{Anber:2018ohz}
M.~M.~Anber and B.~J.~Kolligs,
\emph{Entanglement entropy, dualities, and deconfinement in gauge theories},
JHEP {\bf 1808} (2018) 175
\arxiv{1804.01956}{hep-th}.
  


\bibitem{Cella:1993ic}
G.~Cella, G.~Curci, R.~Tripiccione and A.~Vicere,
\emph{Scaling, asymptotic scaling and Symanzik improvement. Deconfinement temperature in SU(2) pure gauge theory},
Phys.\ Rev.\ D {\bf 49} (1994) 511

\bibitem{GP}
P. Giudice and S. Piemonte, \emph{Improved thermodynamics of SU(2) gauge theory}, Eur. Phys. J. C (2017) 77: 821, \arxiv{1708.01216}{hep-lat}

\bibitem{Lacroix:2014jpa}
G.~Lacroix, C.~Semay and F.~Buisseret,
\emph{The deconfined phase of ${\cal N}=1$ SUSY Yang-Mills: bound states and the equation of state},
\arxiv{1408.4979}{hep-th}.

\bibitem{Witten:1997ep}
E.~Witten,
\emph{Branes and the dynamics of QCD},
Nucl.\ Phys.\ B {\bf 507} (1997) 658
\arxivold{hep-th/9706109}.

\bibitem{Campos:1998db}
A.~Campos, K.~Holland and U.~J.~Wiese,
\emph{Complete wetting in supersymmetric QCD or why QCD strings can end on domain walls},
Phys.\ Rev.\ Lett.\  {\bf 81} (1998) 2420
\arxivold{hep-th/9805086}.















\end{thebibliography}
\end{document}